\documentclass[useAMS,usenatbib,usegraphicx]{mn2e}
\usepackage{bm,amssymb}
\usepackage{times}
\newcommand{\bc}{\begin{center}}
\newcommand{\ec}{\end{center}}
\newcommand{\be}{\begin{equation}}
\newcommand{\ee}{\end{equation}}
\newcommand{\ba}{\begin{array}}
\newcommand{\ea}{\end{array}}
\newcommand{\bea}{\begin{eqnarray}}
\newcommand{\eea}{\end{eqnarray}}

\def\ga{\mathrel{\mathchoice {\vcenter{\offinterlineskip\halign{\hfil
$\displaystyle##$\hfil\cr>\cr\sim\cr}}}
{\vcenter{\offinterlineskip\halign{\hfil$\textstyle##$\hfil\cr
>\cr\sim\cr}}}
{\vcenter{\offinterlineskip\halign{\hfil$\scriptstyle##$\hfil\cr
>\cr\sim\cr}}}
{\vcenter{\offinterlineskip\halign{\hfil$\scriptscriptstyle##$\hfil\cr
>\cr\sim\cr}}}}}
\def\la{\mathrel{\mathchoice {\vcenter{\offinterlineskip\halign{\hfil
$\displaystyle##$\hfil\cr<\cr\sim\cr}}}
{\vcenter{\offinterlineskip\halign{\hfil$\textstyle##$\hfil\cr
<\cr\sim\cr}}}
{\vcenter{\offinterlineskip\halign{\hfil$\scriptstyle##$\hfil\cr
<\cr\sim\cr}}}
{\vcenter{\offinterlineskip\halign{\hfil$\scriptscriptstyle##$\hfil\cr
<\cr\sim\cr}}}}}
\def\mkm{{\mu}\rm{m}}

\def\is{interstellar }
\def\degr{\hbox{$^\circ$}}

\title[Polarization curve and grain growth]
{Effects of grain growth on the interstellar polarization curve}
\author[Voshchinnikov and Hirashita]{Nikolai V. Voshchinnikov$^1$\thanks{E-mail:
    n$_{.}$voshchinnikov@spbu.ru}
and Hiroyuki Hirashita$^2$ \\
$^1$Sobolev Astronomical Institute, St. Petersburg University,
Universitetskii prosp., 28, St. Petersburg 198504, Russia\\
$^2$Institute of Astronomy and Astrophysics, Academia Sinica,
P.O. Box 23-141, Taipei 10617, Taiwan}

\date{Accepted 2014 August 20; Received 2014 August 20; in original form 2014 July 21}

\pagerange{\pageref{firstpage}--\pageref{lastpage}} \pubyear{2014}

\begin{document}
\label{firstpage}
\maketitle

\begin{abstract}
We apply the time evolution of grain size distributions
by accretion and coagulation found
in our previous work
to the modelling of the wavelength dependence of \is linear polarization.
We especially focus on the parameters of the Serkowski curve $K$ and $\lambda_{\max}$
characterizing the width and the maximum wavelength of this curve, respectively.
We use aligned silicate and non-aligned carbonaceous spheroidal
particles with different aspect ratios $a/b$.
The imperfect alignment of grains with sizes larger than
a cut-off size  $r_{V,\rm cut}$ is considered.
{\rm We find that the evolutionary effects on the polarization curve
are negligible in the original model
with commonly used material parameters
(hydrogen number density $n_\mathrm{H}=10^3$ cm$^{-3}$, gas
temperature $T_\mathrm{gas}=10$~K, and the sticking probability for
accretion $S_\mathrm{acc}=0.3$).}
Therefore, we apply the tuned model
where the coagulation threshold of silicate is removed.
{\rm In this model, $\lambda_{\max}$
displaces to the longer wavelengths
and the polarization curve becomes wider ($K$ reduces) on time-scales
$\sim (30 - 50) (n_\mathrm{H}/10^3 \mathrm{cm}^{-3})^{-1}$ Myr.
The tuned models at $T \la 30 (n_\mathrm{H}/10^3 \mathrm{cm}^{-3})^{-1} $~Myr
and different values of the parameters $r_{V,\rm cut}$
can also explain the observed trend between $K$ and $\lambda_{\max}$.}
It is significant that the evolutionary effect appears in the perpendicular
direction to the effect of $r_{V,\rm cut}$ on the
$K$ -- $\lambda_{\max}$ diagram.
Very narrow polarization curves can be reproduced if we change
the type of particles (prolate/oblate) and/or vary $a/b$.
\end{abstract}


\section{Introduction}

Interstellar dust plays a crucial role in the evolution of galaxies
\citep{dwek98,zhukovska08,draine09,inoue11,asano14}.
Traditionally, dust models are
tested using 
\is extinction curves. 
As a result, the constraints on the grain size distribution
(e.g. \citealt{mrn,zubko96,weingartner01})
and evolution of grain materials
(e.g. \citealt{jdw90,lig97,cc10,jones13,mulas13,cc14,kohler14}) can be obtained.

Less popular modelling of \is polarization makes it possible to investigate
not only interstellar dust but also interstellar magnetic fields.
The wavelength dependence of polarization 
is used for estimates of  grain size and composition
(\citealt{mathis86}; \citealt{km95}; \citealt{voshchinnikov13})
while the polarizing efficiency 
and distribution of 
position angles give information about grain alignment and magnetic field
(\citealt{fosetal02}; \citealt{wetal08};  \citealt{vd08}; \citealt{reissl14};
see also \citealt{and12} and  \citealt{voshchinnikov12} for recent reviews).

{\rm In this paper, we examine the variation of \is polarization,
using our recent results of grain size distribution (\citealt{hv14}) and
optical constants of grain materials  (\citealt{jones13}).}

\citet{jones13} proposed a new two-material \is dust model, consisting
of amorphous silicate and hydrocarbon dust. In this model,
the visual -- near infrared (IR) extinction
is mainly produced by large forsterite-type silicate grains with metallic
iron nano-particle inclusions (a-Sil$_{\rm Fe}$) and large
aliphatic-type carbonaceous grains (a-C(:H)). The typical
radii of grains are 0.1~--~0.2~$\mkm$. During the life-cycle
silicate and carbonaceous particles can accumulate 
the thin mantles consisting of aromatic-type carbon (a-C).

\citet[ hereafter HV14]{hv14} have
investigated the time evolution of grain size distribution
due to the accretion and coagulation in an interstellar cloud
and examined whether dust grains
processed by these mechanisms 
can explain observed
variation of extinction curves  in the Milky Way.
It was found that, if we consider accretion and coagulation with commonly
used material parameters, the model fails to explain
the Milky Way extinction curves.
This discrepancy was resolved by adopting
a `tuned' model, in which coagulation of carbonaceous dust
is less efficient  and that of silicate is
more efficient with the coagulation threshold being removed.
The tuned model is also consistent with the relation between
silicon depletion (an indicator of accretion) and $R_V$ (the ratio of total
to selective extinction, an indicator of grain growth), and
the correlation between ultraviolet (UV) slope $c_2$ in the parametric fit
of extinction curve according to \citet{fitzpatrick07} and $R_V$.

Here, we take the size distributions obtained by HV14
as a starting point for our analysis of the time evolution of \is
polarization curve. We use the model of spheroidal particles
that was earlier applied to interpret the \is extinction and
polarization
(\citealt{vd08}; \citealt{dvi10}; \citealt{voshchinnikov13}).

The paper is organized as follows.
The  description of observations and the model is given in
Section~\ref{ispol}. Section~\ref{res} contains the results of
the modelling of wavelength dependence of polarization $P(\lambda)$
and the relation between the width of the curve $P(\lambda)$ and the
position of the maximum polarization $\lambda_{\rm max}$.
Sections~\ref{disc} and \ref{concl} present the discussion of results
and conclusions.

\section{Interstellar polarization}
\label{ispol}

Interstellar linear polarization is caused by the linear dichroism of
the interstellar medium due to the presence of non-spherical aligned
grains. Dust grains must have sizes close to the wavelength
of the incident radiation and specific magnetic properties  to efficiently
interact with the interstellar magnetic field.

\subsection{Observations: Serkowski curve}
\label{obs}

The wavelength dependence of polarization $P(\lambda)$ in the visible part of
spectrum is described by an empirical formula suggested by
\citet{serk73}
\be
P(\lambda)/P_{\max} = \exp [-K \ln^2 (\lambda_{\max}/\lambda)].
\label{serkk}
\ee
This formula has three parameters: the maximum degree of polarization $P_{\max}$,
the wavelength corresponding to it $\lambda_{\max}$ and
the coefficient $K$ characterizing the width of the Serkowski curve.
The values of $P_{\max}$ in the diffuse interstellar medium usually
do not exceed 10\%, and the mean value of $\lambda_{\max}$ is 0.55\,$\mkm$
(\citealt{smf75}).

The parameter $K$ determines the half-width
of the normalized \is linear polarization curve
\be
W = \lambda_{\rm max}/\lambda_{-} -
 \lambda_{\rm max}/\lambda_{+},
\ee
where
$\lambda _{-} < \lambda _{\max } < \lambda _{+}$ and 
$P(\lambda _{+}) = P(\lambda _{-}) = P_{\max }/2$.
The relation between $W$ and $K$ is as follows:
\be
W = \exp [(\ln 2/K)^{1/2}] - \exp [-(\ln 2/K)^{1/2}].
\label{wk}
\ee
Treating $K$ as a third free parameter of the
Serkowski curve, \citet{wetal92} fit the relation
between $K$ and $\lambda_{\max}$ in the Milky Way as
\be
K = (1.66 \pm 0.09) \lambda_{\max} + (0.01 \pm 0.05).
\label{k92}
\ee

\subsection{Modelling}
\label{mod}

{\rm  Interpretation of the wavelength dependence of \is polarization
includes calculations of the extinction cross sections of
rotating partially aligned non-spherical particles.
In early studies, an unphysical model of infinitely long cylinders
was applied.
More realistic is the spheroidal model of grains with
the shape of these axisymmetric particles being
characterized by the only parameter ---
the ratio of the major to minor semi-axis $a/b$.
This  model is particularly promising for interpretation of the \is
polarization and extinction data
(see \citealt{voshchinnikov12} for review).}

We represent the \is dust grains by a mixture of silicate and carbonaceous
homogeneous spheroids of different sizes and orientations.
A solution to the light scattering problem for such particles
has been given by \citet{vf93}.

The linear polarization of unpolarized stellar radiation
produced by aligned rotating spheroidal particles after the passage
through a dust cloud with the uniform magnetic field\footnote{The angle
between the line of sight and the magnetic field
is denoted by $\Omega$ ($0\degr \leq \Omega \leq 90\degr$).}
is
\bea
 P(\lambda)= \sum_j \int\limits_0^D \int\limits_{r_{V,\min,j}}^{r_{V,\max,j}}
    \overline{C}_{{\rm pol},j}(m_{\lambda,j},a_j/b_j,r_{V},\lambda)\,
 \nonumber \\ \times
    n_{j}(r_{V})\, dr_{V} \,dl \times 100\,\%\,,
\eea
\bea
 \overline{C}_{{\rm pol},j}
   = {\frac{2}{\pi^2}}
   {\int\limits_{0}^{\pi/2}}{\int\limits_{0}^{\pi}}{\int\limits_{0}^{\pi/2}}
   \frac{1}{2}
   (C^{\rm TM}_{{\rm ext},j}-C^{\rm TE}_{{\rm ext},j}) \,
 \nonumber \\ \times
    f_j(\xi, \beta, ...) \, \cos 2{\psi} \, d{\varphi}\, d{\omega}\, d{\beta} \,,
\label{cpol}
\eea
where $D$ is the distance to the star, 
$\lambda$ the wavelength,
$m_{\lambda,j}$, $a_j/b_j$ è $n_{j}(r_{V})$ are
the refractive index, aspect ratio and size distribution
of spheroidal particles of the $j$th kind
($j=$Si for silicate and $j=$C for carbonaceous dust, respectively),
$r_{V}$ is the radius of a sphere whose volume is equal to that of the spheroid
(for prolate particles, $r_{V} = \sqrt[3]{a b^2}$, and
for oblate ones, $r_{V} = \sqrt[3]{a^2 b}$),
${r_{V,\min,j}}$ and ${r_{V,\max,j}}$ are the minimum and maximum radii,
respectively, and
$C^{\rm TM, \, TE}_{{\rm ext},j}$
the extinction cross-sections for two polarization modes
depending on the particle orientation relative to the electric vector of
incident radiation (\citealt{bh83}).
The angle $\psi$ can be expressed through $\varphi, \omega, \beta, \Omega$
(see the definitions of the angles and relations between them,
e.g., in \citealt{dvi10}), and finally
${f}_j(\xi, \beta, ...)$ is the distribution of the particles of the $j$th
kind over orientations.   

\subsubsection{Size distribution}\label{size}

\begin{figure*}
\resizebox{8.5cm}{!}{\includegraphics{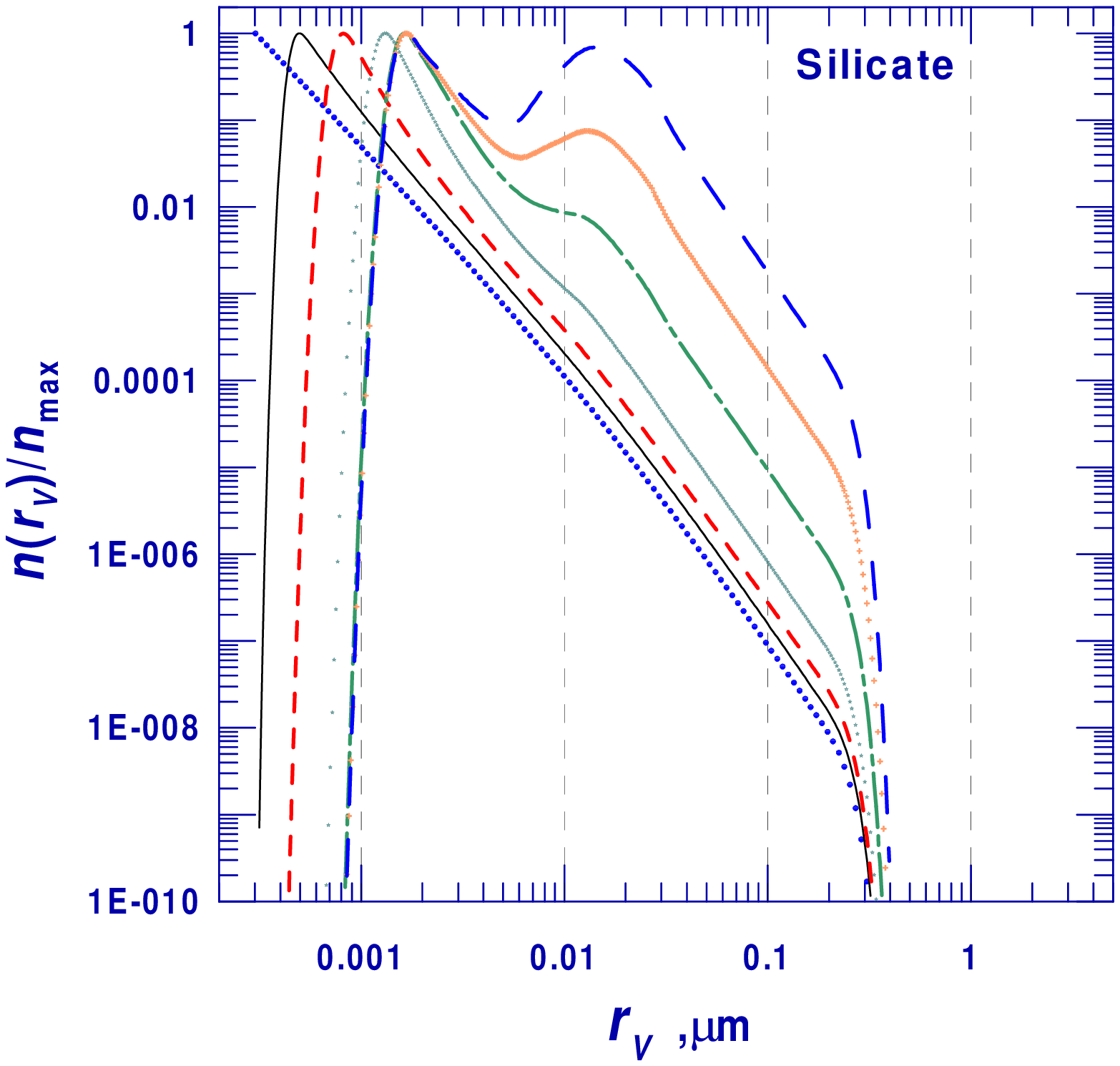}}
\resizebox{8.5cm}{!}{\includegraphics{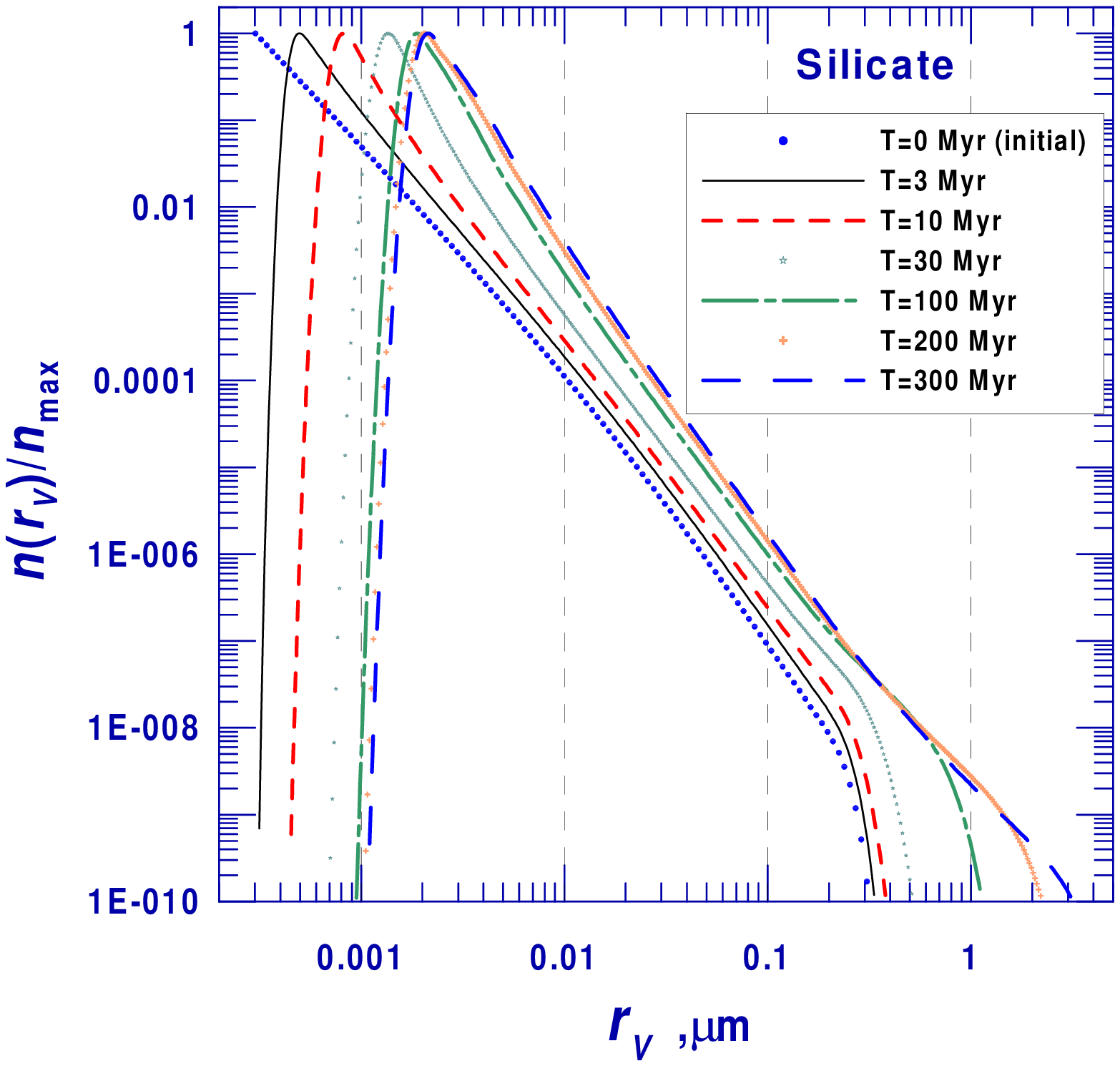}}
\caption{Evolution of grain size distribution for
silicate dust. The grain size distribution is normalized to the maximum.
Effects of evolution of grain size distribution
are shown for original (left panel)  and tuned (right panel) models
at $T=3$, 10, 30, 100, 200 and 300 Myr.
The initial condition is shown by the dotted line.
}
\label{nrsi}
\end{figure*}

HV14 calculated the time evolution of grain size distribution by
accretion and coagulation based on the formulation by \citet{hirashita12}.
Silicate and carbonaceous dust were treated as separate grain species.

The growth time-scale of grain radius for accretion is determined by
the gas-phase abundance of the key elements Si and C
(assumed to be proportional to the metallicity $Z$),
hydrogen number  ($n_\mathrm{H}$) density, gas
temperature ($T_\mathrm{gas}$) and the sticking probability for
accretion ($S_\mathrm{acc}$).
We adopt the following values  for quantities:
$Z=\mathrm{Z}_{\sun}$, $n_\mathrm{H}=10^3$ cm$^{-3}$,
$T_\mathrm{gas}=10$~K, and $S_\mathrm{acc}=0.3$.

HV14 considered thermal (Brownian)
motion and turbulent motion as a function of grain mass
and solved a discretized coagulation equation
(see \citealt{hirashita12} for details).
The coagulation was assumed to occur if the
relative velocity was below the coagulation
threshold given by \citet{hirashita09}.
It was also adopted that the sticking coefficient
for coagulation $S_\mathrm{coag}$ was 1.

As an initial grain size distribution (at time $T=0$),
HV14 choose the size distributions of silicate and carbonaceous dust that
fits the mean Milky Way extinction curve for $R_V=3.1$
\citep{weingartner01}.
The initial size distribution of silicate grains
is shown by the dotted line in Fig.~\ref{nrsi}.

Using the size distributions obtained for silicate and carbonaceous dust,
HV14  calculated the evolution of extinction curve and
found that the prominent variation of extinction
caused by increasing accretion occurs at $< 30$ Myr.
After that, the evolution of extinction curve is driven only by
coagulation, which flattens the extinction curve and makes
$R_V$ larger.
Observations indicate that even for large $R_V \approx 4.4$,
the carbon bump around 0.22 $\micron$ is clear and
the UV extinction rises with a positive curvature.
However, the theoretical predictions show that at large $R_V$
(i.e.\ after significant coagulation at $T>100$ Myr),
the carbon bump disappears. This implies that coagulation
is in reality not so efficient as assumed in the model
for carbonaceous dust. There is another discrepancy:
the observed $A_V/N_\mathrm{H}$ tends to decrease
drastically as $R_V$ increases,
while it does not decrease significantly after pronounced
coagulation at $T>100$ Myr.
This is
because silicate stops to coagulate at $r_V\sim 0.03~\micron$,
which is still too small to affect the UV opacity.

In summary, the observational data indicate
(i) that carbonaceous dust should be more inefficient in
coagulation than assumed, and (ii) that silicate should grow
beyond $r_V\sim 0.03~\micron$ (the growth is limited
by the coagulation threshold velocity).
{\rm Therefore, for a better fit the observational data,
HV14  also tried models with
$S_\mathrm{coag}<1$ for carbonaceous dust and
no coagulation threshold for silicate dust as
a tuned model.
Such a tuning is acceptable if we consider uncertainties in
grain properties. The removal of the coagulation threshold
may be justified if the grains have fluffy structures that
absorb the collision energy efficiently, and/or are coated with sticky
materials such as water ice (\citealt{ormel09,hirashita13}).}

In the tuned model, the coagulation efficiency was decreased
by a factor of 2 by adopting
$S_\mathrm{coag}=0.5$ for carbonaceous dust, while
$S_\mathrm{coag}=1$ for silicate was kept.
Silicate grains were assumed to coagulate whenever they collide
(no coagulation threshold).
The accretion efficiency ($S_\mathrm{acc}=0.3$)
was not changed for both silicate and
carbonaceous dust to minimize the fine
tuning. This is called `tuned model', while
we call the original model with $S_\mathrm{coag}=1$
and the coagulation threshold `original model
without tuning'.

The evolution of silicate grain size distribution for the tuned
model is shown in Fig.~\ref{nrsi} (right panel).
Compared with the dust distributions in
the original model without tuning (Fig.~\ref{nrsi}, left panel),
the grains grow further, even beyond 1~$\micron$
at $>100$ Myr.
As shown later, silicate grains in the large-size tail
of size distribution are of crucial importance in explanation of
observed polarization.

We use size distribution of silicate and carbonaceous grains obtained by
HV14 but replace spherical particles by spheroids of the same mass.
{\rm The uncertainties of grain shape to spheroidal has only a
minor influence on 
accretion and coagulation compared with the change of
$S_\mathrm{acc}$ and $S_\mathrm{coag}$.
Therefore, that we simply use HV14's results
with an assumption of $a=r_v$.}

\subsubsection{Materials}

\citet{jones12a,jones12b,jones12c} published
a comprehensive analysis of the compositional properties of hydrogenated
amorphous carbons.
It includes the consideration of the evolution of carbon solids from
hydrogen rich aliphatic grains (a-C:H, sp$^3$ bonded) to
hydrogen poor aromatic grains (a-C, sp$^2$ bonded), calculations of the complex
refractive indexes as a function of the material band gap, and
investigation of size-dependent properties.
Later, \citet{jones13} found refractive indexes of amorphous silicates
using the mixture of amorphous forsterite with different amount of metallic
iron (a-Sil$_{\rm Fe}$) and
\citet{kohler14} added iron sulfide in this mixture.

In modelling, we choose the refractive index
'si\-li\-ca\-te$\_$FoFe10.RFI' for amorphous silicates with 10\% volume
fraction of Fe (equivalent to $\sim$70\% of the cosmic Fe) and
the data for an a-C(:H) material with a band gap $E_g=2.5$~eV.
The optical constants were taken from \citet{jones13} and \citet{jones12b}
for silicate and carbon, respectively.
The chosen carbon material seems to represent the bulk of the
carbonaceous dust
formed around evolved stars (A. P. Jones, private communication).
We ignore the possible evolutionary sequence for the silicates
because of the large vagueness of this problem and
the size dependence of refractive indexes since it is unimportant
in the visual part of spectrum.
Note that the used materials are different from those in HV14.
However, this fact as well as the replacement of spherical particles
by spheroidal ones are of no significance for normalized extinction
curves, e.g., the difference in  $R_V$ does not exceed 0.1 -- 0.2.

\subsubsection{Alignment}

We assume that the spheroidal grains are partly aligned so that
their major axes rotate in a plane ($\varphi$ is the angle of rotation)
and their angular momentum {\rm J} precesses around the direction of
the magnetic field ($\omega$ is the precession angle,
$\beta$ the opening angle of the precession cone). 
Such alignment is called the imperfect Davis--Greenstein (IDG) alignment.
 It is described by the distribution function ${f}_{{\rm IDG},j}(\xi, \beta)$
depending only on the orientation parameter $\xi$ and the angle $\beta$
for particles of the $j$th kind {\rm (\citealt{hg80})}:
\be
{f}_{{\rm IDG},j}(\xi, \beta) = \frac{\xi \sin \beta}{(\xi^2 \cos^2 \beta + \sin^2 \beta)^{3/2}}.
\label{idg}
\ee
The parameter $\xi$ depends on the particle size $r_V$,
the imaginary part of the magnetic susceptibility of a dust grain
$\chi''=\varkappa \omega_{\rm d} /T_{\rm d}$,
where $\omega_{\rm d}$ is the angular velocity of the particle,
hydrogen number density $n_{\rm H}$, magnetic field strength $B$, and
temperatures of dust $T_{\rm d}$ and gas $T_{\rm gas}$ as
\be
\xi^2 = \frac{r_V +\delta_{0,\,j}^{\rm IDG} (T_{\rm d}/T_{\rm gas})}{r_V
+ \delta_{0,\,j}^{\rm IDG}},
\label{xi} \ee
where
\be
\delta_{0,\,j}^{\rm IDG} = 8.23\,10^{23} \frac{\varkappa B^2}{n_{\rm H} T_{\rm gas}^{1/2} T_{\rm d}}\,\mkm.
\label{delta} \ee

In the IDG mechanism, smaller grains which are powerful polarizers
are aligned better than larger grains (see Eq.~(\ref{xi})).
In order to reduce the polarizing efficiency below the observed values,
it is possible to increase the minimum cut-off in grain size
distribution ${r_{V,\min,j}}$ (e.g. \citealt{dvi10}) or
to assume that small particles are randomly oriented (e.g. \citealt{df09}).

Because the grain size distribution is fixed in our model
(see Section~\ref{size}) we suggest the following modified IDG alignment
function with reduced alignment of small grains (see also \citealt{mathis86}):
\be
 {f}_j(\xi,\beta,...) =
\left [  1- \exp (-r_V/r_{V,\rm cut})^3 \right ] \, \times \,
 {f}_{{\rm IDG},j}(\xi, \beta) \,,
\label{fj}
\ee
where $r_{V,\rm cut}$ is a cut-off parameter.
The alignment function (\ref{fj}) gives a smooth
switch from non-aligned to aligned grains in comparison with
abrupt jump  suggested by \citet{df09}.

 Though the silicate and carbonaceous particles comparably
contribute to extinction, the polarization is assumed to be produced only by 
silicate particles. Such an assumption has been earlier done by
 \citet{chkru83} and \cite{mathis86},
and recently has got additional support in the work of \citet{vhpd12}
who found a correlation between the observed \is polarization degree and
the abundance of silicon in dust grains.
Polarization in  IR features also
supports the idea of separate populations of polarizing
(silicate) and non-polarizing (carbonaceous) grains. This follows
from the observed polarization of silicate features at
10\,$\,\mkm$ and 18\,$\,\mkm$ and the lack of polarization
in the 3.4\,$\,\mkm$ hydrocarbon feature (\citealt{ha04,chair06};
see also discussion in \citealt{li14}).


\section{Results}\label{res}

We made calculations for prolate and oblate homogeneous spheroids
consisting of silicate and amorphous carbon. The particles of 59
sizes in the range from $r_{V, \min}=0.001 \,\mkm$ to $r_{V, \max} = 1 \,\mkm$
were utilized.
The extinction and polarization curves have been calculated for 77
wavelengths in the range from $\lambda=0.2\,\mkm$ to $\lambda=5\,\mkm$.
We next determine parameters of the Serkowski curve $P_{\max}$,
$\lambda_{\max}$, and $K$.
 In this paper, we focus on
the normalized curves $P(\lambda)/P_{\max}$ and
the relation between the width of
the polarization curve and the position of its maximum. 
 These characteristics of the Serkowski curve are mainly determined
by the size distribution $n_{\rm Si}(r_{V})$
and the cut-off parameter $r_{V,\rm cut}$
of silicate particles
and weakly depend on the degree ($\delta_{0}^{\rm IDG}$)
and direction ($\Omega$) of the particle orientation.
Therefore, we fix the alignment parameters:
$\delta_{0,\,{\rm Si}}^{\rm IDG}=0.5\,\mkm$, $\delta_{0,\,{\rm C}}^{\rm IDG}=0.01\,\mkm$, and
$\Omega=60\deg$.


\subsection{Wavelength dependence}\label{subsec:wave}
\subsubsection{Linear polarization}

\begin{figure}
\centerline{
\resizebox{\hsize}{!}{\includegraphics{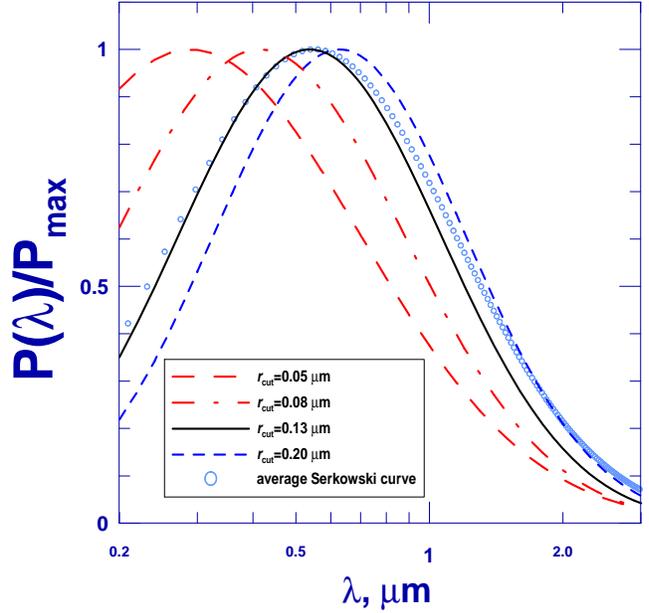}}}
\caption{Normalized wavelength dependence of linear polarization
for the initial size distribution ($T=0$~Myr, see Fig.~\ref{nrsi}).
The average observational Serkowski curve ($\lambda_{\max}=0.55\,\mu$m,
$K=0.92$) is plotted using the open circles.
The theoretical curves were calculated for
prolate spheroids with $a/b=3$.
The characteristics of theoretical curves are:
$r_{V,\,\rm cut}=0.05\,\mu$m, $\lambda_{\max}=0.289\,\mu$m, $K=0.64$;
$r_{V,\,\rm cut}=0.08\,\mu$m, $\lambda_{\max}=0.415\,\mu$m, $K=0.88$;
$r_{V,\,\rm cut}=0.13\,\mu$m, $\lambda_{\max}=0.538\,\mu$m, $K=1.07$ and
$r_{V,\,\rm cut}=0.20\,\mu$m, $\lambda_{\max}=0.628\,\mu$m, $K=1.16$.
\label{cut}}
\end{figure}

\begin{figure*}
\resizebox{8.5cm}{!}{\includegraphics{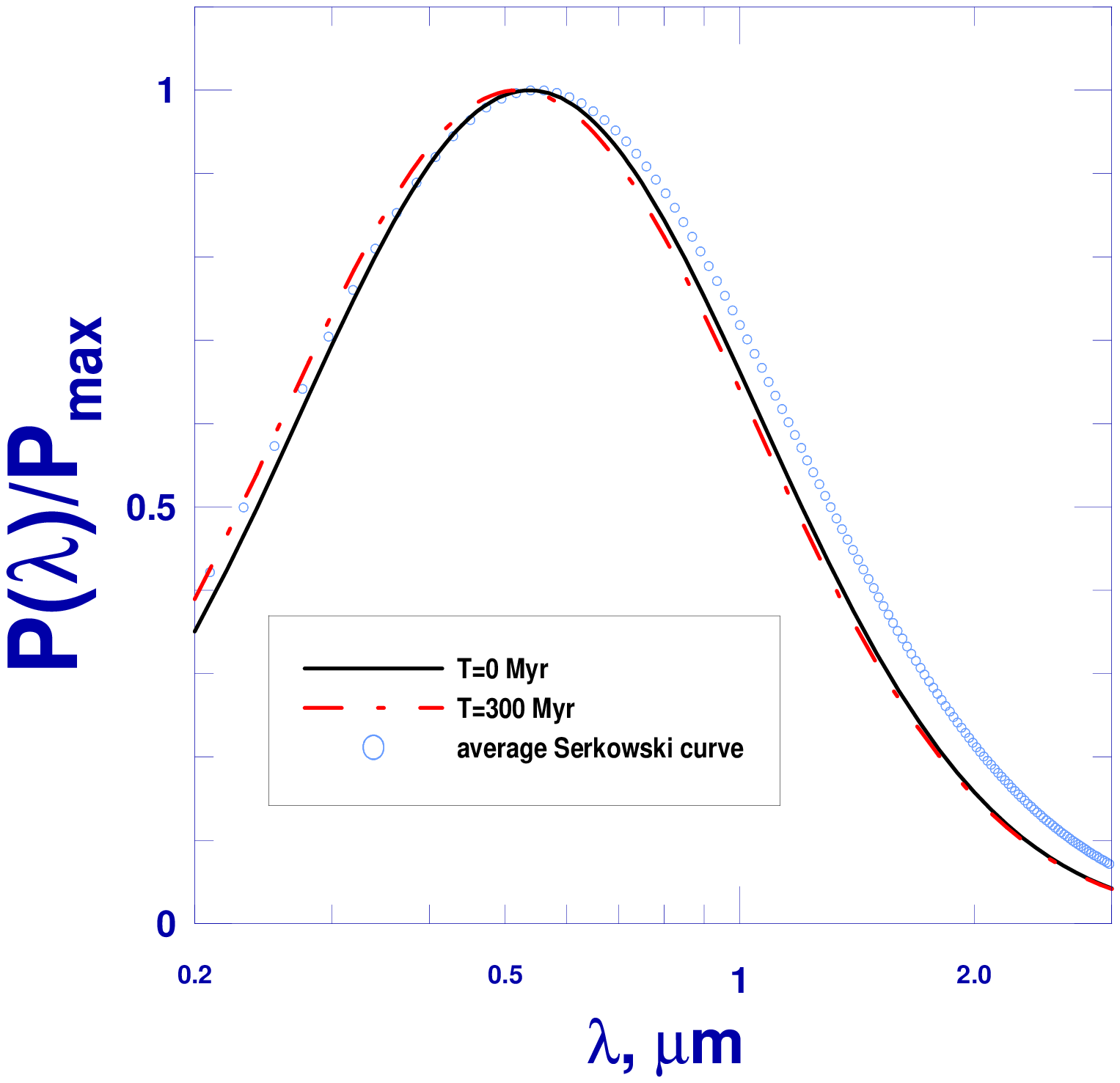}}
\resizebox{8.5cm}{!}{\includegraphics{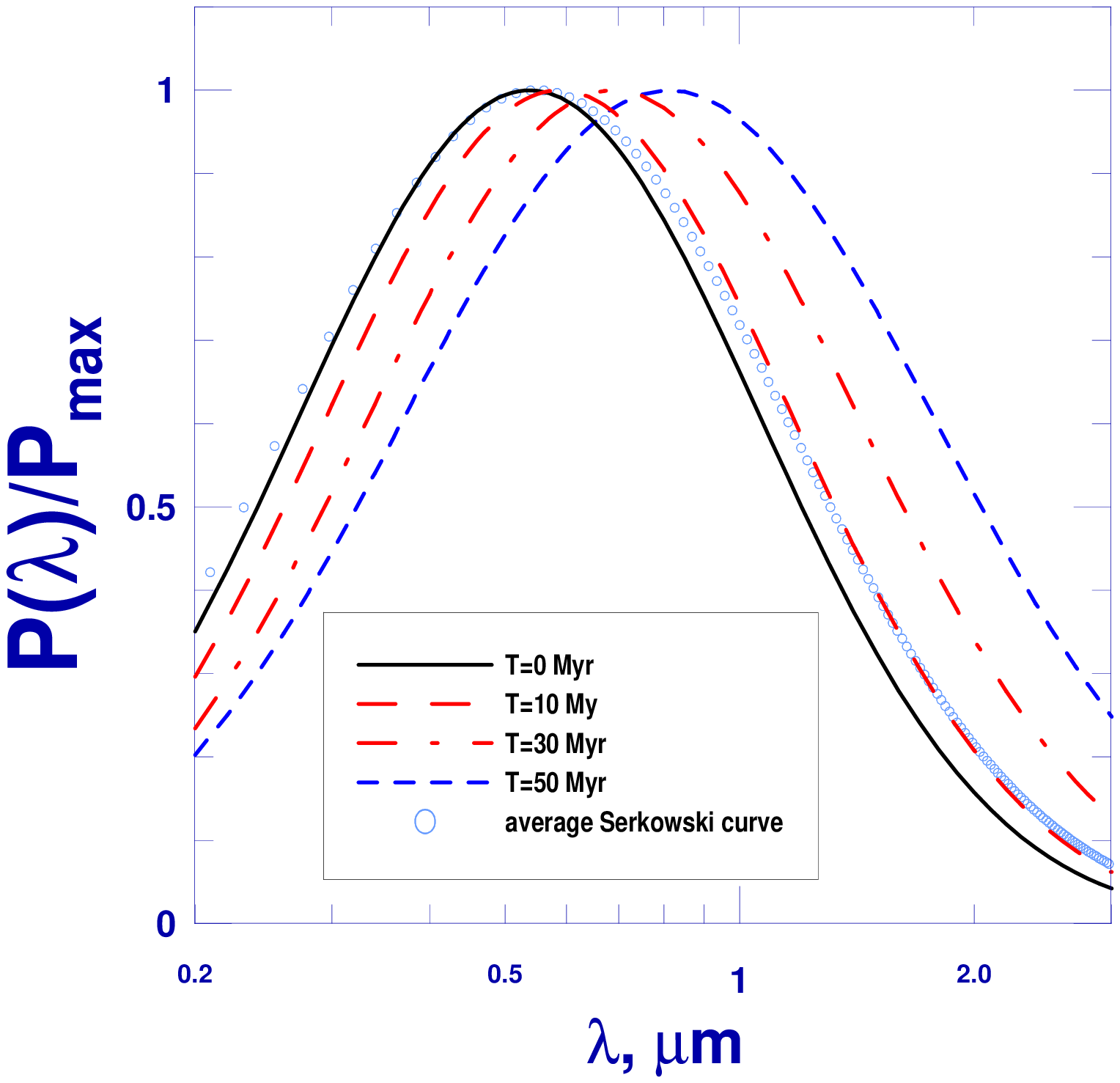}}
\caption{Normalized wavelength dependence of interstellar
linear polarization. 
Effects of evolution of grain size distribution on curves $P(\lambda)$
are shown for the original model without tuning (left) and
the tuned (right) model.
The average observational Serkowski curve ($\lambda_{\max}=0.55\,\mu$m,
$K=0.92$) is plotted using the open circles.
The theoretical curves were calculated for the model:
prolate spheroids, $a/b=3$, $r_{V,\,\rm cut}=0.13\,\mu$m.
The parameter $r_{V,\,\rm cut}$ for initial curve was chosen in such a
manner in order to be closer to observational curve.
The characteristics of theoretical curves are:
$T=0$ Myr,   $\lambda_{\max}=0.538\,\mu$m, $K=1.07$;
original model:
$T=300$ Myr, $\lambda_{\max}=0.519\,\mu$m, $K=1.04$;
tuned model:
$T=10$ Myr, $\lambda_{\max}=0.588\,\mu$m, $K=1.05$;
$T=30$ Myr, $\lambda_{\max}=0.689\,\mu$m, $K=0.95$;
$T=50$ Myr, $\lambda_{\max}=0.812\,\mu$m, $K=0.81$.}
\label{plam}
\end{figure*}

First of all, we calculated the polarization curve
$P(\lambda)/P_{\max}$ for the initial size distribution of silicate and
carbonaceous dust. If we assume that grains of all sizes are
aligned according to IDG mechanism ($r_{V,\,\rm cut}=0$ in Eq.~(\ref{fj})),
the polarization peaks at far-UV ($\lambda_{\max}< 0.2\,\mu$m).
In order to fit the average observational Serkowski curve
($\lambda_{\max}=0.55\,\mu$m and $K=0.92$ according to Eq.~(\ref{k92})),
we increase parameter $r_{V,\,\rm cut}$.
Figure~\ref{cut} shows how the variations of $r_{V,\,\rm cut}$ influence
the polarization curve. It is clearly seen that both
$\lambda_{\max}$ and $K$ grow with growing the cut-off parameter,
i.e. for larger values of $r_{V,\,\rm cut}$,
the curve $P(\lambda)$ becomes  narrower and its maximum shifts
to longer wavelengths (see Section~\ref{klm} for more discussion
the relation between $K$ and $\lambda_{\max}$). A plausible result
would be achieved for $r_{V,\,\rm cut}=0.13\,\mu$m (Fig.~\ref{cut}).

We choose the model with  $r_{V,\,\rm cut}=0.13\,\mu$m presented
in Fig.~\ref{cut} as the basic one. For this model and the assumed values
of $\delta_{0}^{\rm IDG}$ and $\Omega$,
the ratio of total extinction to selective one $R_V=3.39$
and the polarizing efficiency of the \is medium
$P_{\max}/A_V=1.54$\,\%/mag.
(usually $P_{\max}/A_V \la 3$\,\%/mag. for \is polarization,
\citealt{smf75}).\footnote{\rm Note that the normalized curves
$P(\lambda)/P_{\max}$ quite well reproducing observations
can be obtained if we use the model with non-aligned small grains
and perfectly aligned large grains (perfect Davis--Greenstein alignment).
A sharp or smooth cut-off can be considered; however, in the former case,
the resulting polarizing efficiency $P_{\max}/A_V$ is three times larger
than the observational maximum.}
The average size of polarizing dust grains
b$\langle r_{V,{\rm pol,\,Si}} \rangle = 0.17\,\mkm$.
It can be found using the following expression:
\be
\langle r_{V,{\rm pol,\,Si}} \rangle  = \frac{\displaystyle \int\limits_{r_{_{V},{\rm cut}}}^{r_{_{V},\max}} r_{V} n_{\rm Si}(r_{V}) \,{d}r_{V}}
{\displaystyle \int\limits_{r_{_{V},{\rm cut}}}^{r_{_{V},\max}} n_{\rm Si}(r_{V})\, { d}r_{V}}\,.
\label{rrr}\ee
With increasing  $r_{V,\,\rm cut}$ from $0.05\,\mu$m to $0.20\,\mu$m,
$\lambda_{\max}$ and $K$ grow from $0.289\,\mu$m to $0.628\,\mu$m and
from 0.64 to 1.16, respectively, while
$\langle r_{V,{\rm pol,\,Si}} \rangle$
increases from $0.08\,\mkm$ to $0.23\,\mkm$.

Now we can investigate the changes
in polarization curve owing to the evolution of grain size distribution.
Our results are shown in Fig.~\ref{plam} for original model without
tuning (left panel) and tuned model (right panel).
As is clearly seen from Fig.~\ref{plam}, the evolutionary effects are negligible
in the original model: the difference between initial polarization curve
and that at $T=300$ Myr is very small and comparable to observational errors.
The reason is that the significant changes of grain size occur only
at the smallest sizes for silicate grains which
do not contribute to the observed polarization (Fig.~\ref{nrsi}, left panel)
while the largest size of silicate grains is always $\sim 0.3\,\mkm$.
However, there are time variations in extinction curves,
especially in the UV bump
because the carbonaceous dust can grow beyond $0.3\,\mkm$
(see discussion in HV14).

In the tuned model, the coagulation threshold of silicate is removed,
which leads to the appearance of rather large grains (Fig.~\ref{nrsi}, right panel).
As a consequence, the curves $P(\lambda)$ show a detectable shift
from the initial curve just after $T \ga 10$~Myr (Fig.~\ref{plam}, right
panel). During first stage ($\sim 30 - 50$ Myr), the maximum of polarization
displaces to  longer wavelengths and the polarization curve becomes
wider ($K$ reduces).
Further evolution shows that $\lambda_{\max}$ continues
to grow but the curve $P(\lambda)$ becomes narrower ($K$ increases).
It is evident that the polarization curves obtained
for $T \ga 50$~Myr do not reproduce the available observations
(see Fig.~\ref{k-lms} and discussion in Section~\ref{klm}).

\begin{figure}
\centerline{
\resizebox{\hsize}{!}{\includegraphics{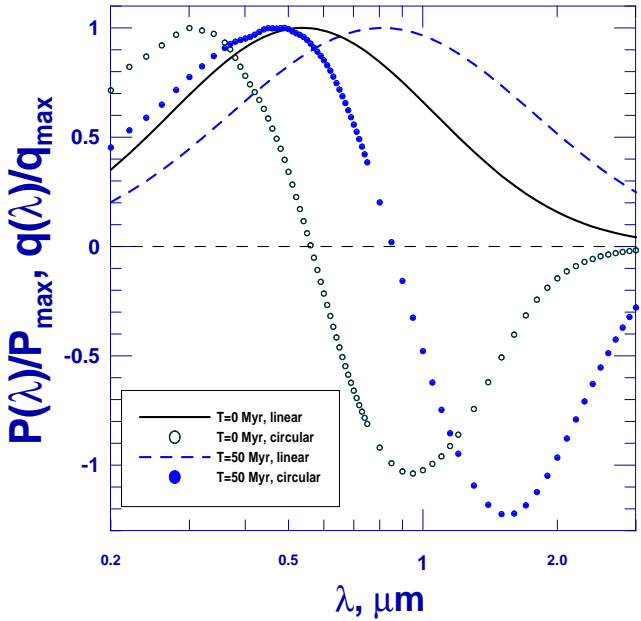}}
}
\caption{Normalized wavelength dependence of interstellar
linear and circular polarization.
The curves $P(\lambda)P_{\max}$ and $q(\lambda)/q_{\max}$
are shown for the initial curve ($T=0$~Myr) and in the case
of tuned model ($T=50$~Myr).}
\label{qpol}
\end{figure}

\subsubsection{Circular polarization}

\begin{figure*}
\centerline{
\resizebox{8.5cm}{!}{\includegraphics{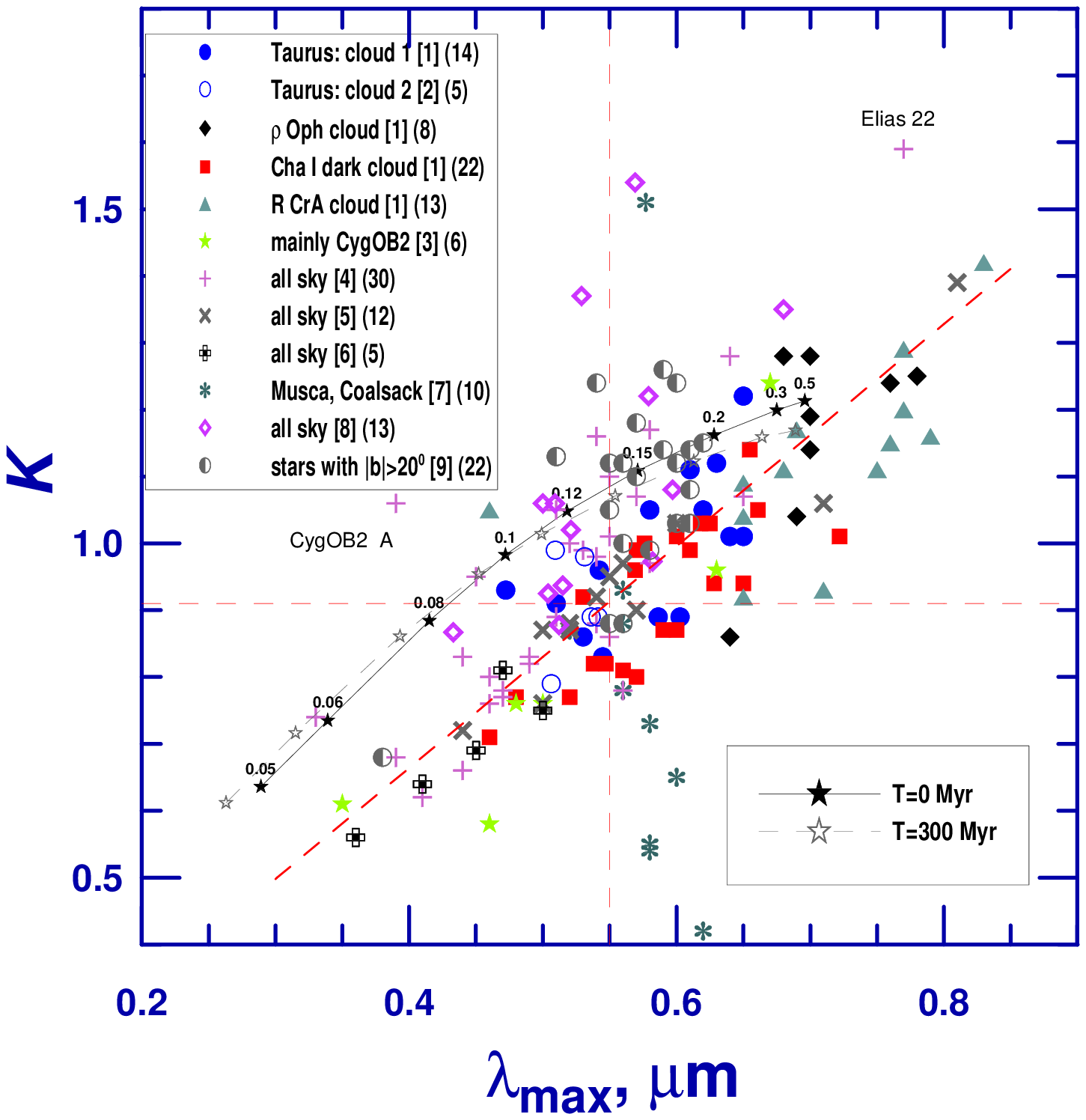}}
\resizebox{8.5cm}{!}{\includegraphics{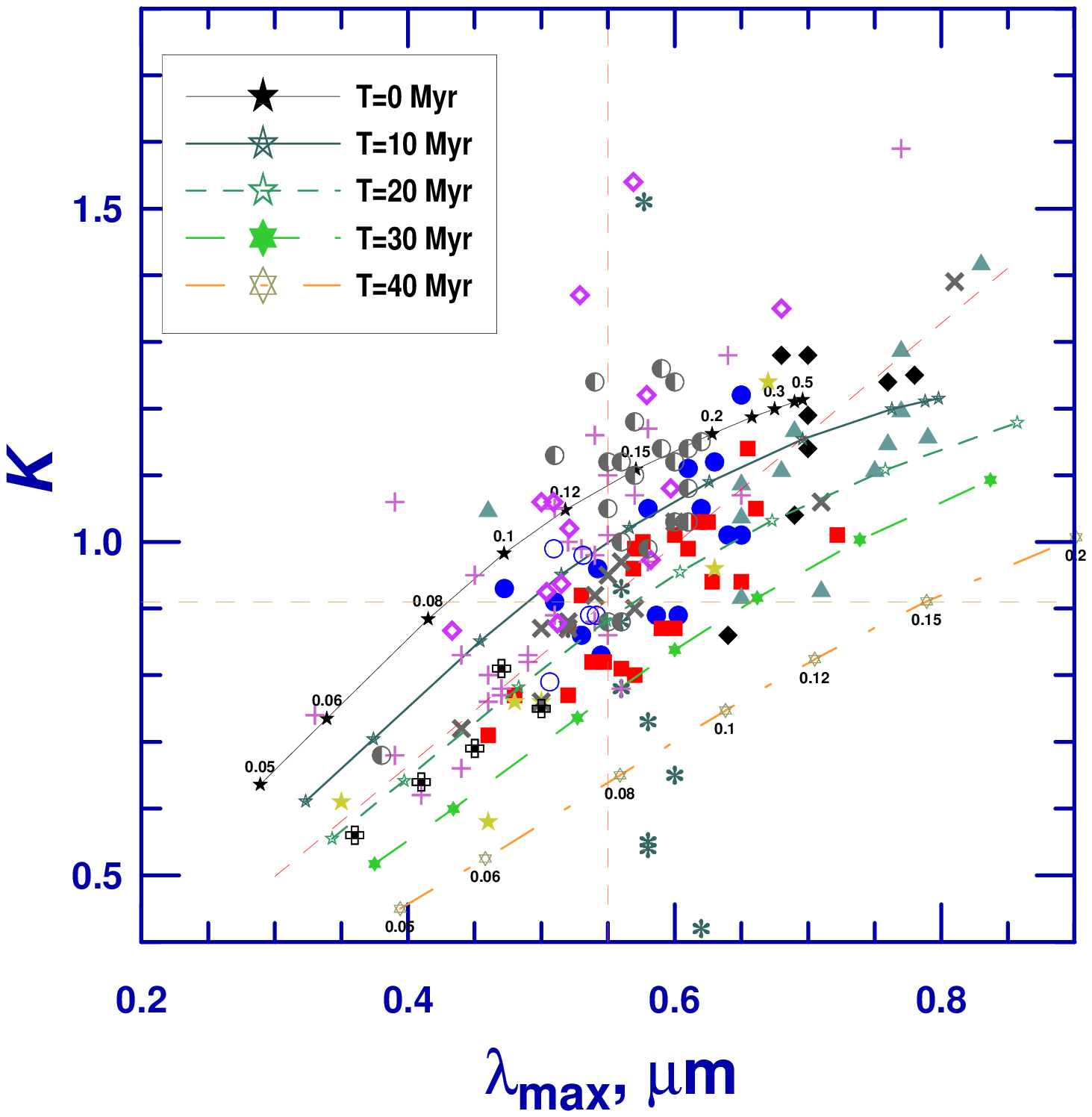}}
}
\caption{Parameter $K$ of Serkowski curve versus
the wavelength $\lambda_{\max}$ of the maximum linear polarization.
The symbols indicate the observational data for 160 stars in the Milky Way.
The sources of data and number of stars are given in the legend
in square brackets and parenthesis, respectively
([1] -- \citealt{voshchinnikov13};
 [2] -- \citealt{wetal01};
 [3] -- \citealt{maretal92};
 [4] -- \citealt{wetal92};
 [5] -- \citealt{wilketal80};
 [6] -- \citealt{wilketal82};
 [7] -- \citealt{ap07};
 [8] -- \citealt{clay95};
 [9] -- \citealt{lar99}).
Stars ($\star$) connected with solid and dashed lines illustrate the results
of modelling for original model without tuning (left panel) and
tuned model (right panel). Model: prolate spheroids with $a/b=3$
are adopted.
The values of $r_{V,\,\rm cut}$ are shown for the initial
size distribution  ($T=0$~Myr) and in the case
of tuned model ($T=40$~Myr).
Horizontal and vertical dashed lines
correspond to the average Serkowski curve ($\lambda_{\max}=0.55\,\mu$m,
$K=0.92$). Inclined dashed line corresponds to the relation
$K = 1.66  \lambda_{\max}$ (see Eq.~(\ref{k92})).}
\label{k-lms}
\end{figure*}

We also calculated the wavelength dependence of \is
circular polarization $q(\lambda)$ for the initial size distribution
and for the tuned model. These curves are plotted in Fig.~\ref{qpol}
together with the corresponding linear polarizations $P(\lambda)$.
As usual, the circular polarization has two extremes
(see e.g. \citealt{v04,Siebenmorgen14}). They shift with time to longer
wavelengths
The wavelength where the curve $q(\lambda)$ changes the sign is close
to the wavelength where the curve $P(\lambda)$ has the maximum
(i.e., $\lambda_{c} \approx \lambda_{\max}$), which, in particular,
confirms the correctness of used data for refractive indexes.

\subsection{Relation between $K$ and $\lambda_{\max}$}
\label{klm}

Now we discuss the models in the context of
the sample of Milky Way polarization curves
with the determined parameters of Serkowski curve collected from the literature.
The starting list contained 57 stars located
in the nearby dark clouds in Taurus and Chamaeleon, around the stars $\rho$ Oph
and R CrA (\citealt{voshchinnikov13}). It was extended up to 160 objects
using the data published by \citet{maretal92}, \citet{wetal92}, \citet{wilketal80},
\citet{wilketal82}, \citet{ap07}, \citet{clay95}, and \citet{lar99}.
The observational data are plotted in Fig.~\ref{k-lms} using different symbols.
The stars are located in different parts of the sky and at different distances.
Although we excluded the stars with rotation of position angle,
perhaps, some stars are observed through two or more clouds.
Sometimes, this may distort the values of $K$ and $\lambda_{\max}$ in comparison
with the single cloud situation (\citealt{clarke84,clarke10}).
Nevertheless, the bulk of observational points concentrates in the
middle part of Fig.~\ref{k-lms} and shows  a clear correlation
between parameters $K$ and $\lambda_{\max}$.
The typical errors are $\sim 0.01\,\mkm$ for $\lambda_{\max}$ and $\sim 0.1$
for $K$. Almost all stars located  above and below the general trend (excluding
may be two objects CygOB2 A and Elias 22) are normal stars. The observational
data for the stars with $K$ and $\lambda_{\max}$ beyond the general trend
as well as the regional variations require more careful analysis
which will be made in a separate paper. Note also that
stars at high galactic latitudes obtained by \citet{lar99}
do not stand out between others.

Theoretical dependence constructed for the
initial model and different $r_{V,\,\rm cut}$
demonstrate the simultaneous growth of $K$ and $\lambda_{\max}$
but passes slightly above the major cluster of observational data.
For the used model (prolate spheroids, $a/b=3$) there is
another problem with the explanation of the observational points with
$\lambda_{\max} \ga 0.7\,\mu$m. This is difficult
even for models with very large values of $r_{V,\,\rm cut}$.
An attempt to remedy the situation using the evolved
size distribution in the original model without tuning
does not meet a success (Fig.~\ref{k-lms}, left panel):
the shift of points is very insignificant.
However, if we apply the tuned model, the major part of observational data
can be fitted. Figure~\ref{k-lms} (right panel) shows that
the tuned models with $T \la 40$~Myr allow simultaneously
to increase the maximum wavelength and the width of polarization curve
($K$ reduces) in comparison with the initial model.


Very narrow polarization curves ($K>1.2$)
as well as observational points located above the  model with $T=0$~Myr
can be reproduced if we change the type of
particles (prolate/oblate) or/and to vary
the particle shape (parameter $a/b$) (Fig.~\ref{k-abt}).
The replacement of prolate spheroids by oblate ones leads to a
growth of both $\lambda_{\max}$ and $K$ (the curves become narrower).
The same dependence takes  place
if we decrease the aspect ratio $a/b$ for prolate
spheroids. The  opposite situation occurs for oblate spheroids:
$\lambda_{\max}$ and $K$ grow if  $a/b$ is increased.

\section{Discussion}\label{disc}

\begin{figure}
\centerline{
\resizebox{\hsize}{!}{\includegraphics{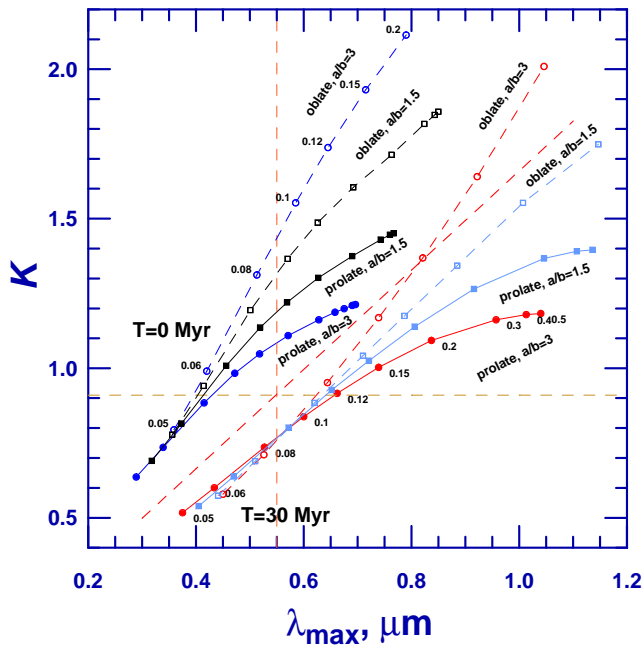}}
}
\caption{Parameter $K$ of Serkowski curve versus
the wavelength $\lambda_{\max}$ of the maximum linear polarization.
The results of modelling for the initial size distribution  ($T=0$~Myr)
and for tuned model with $T=30$~Myr are shown in the case of
prolate and oblate spheroids with  $a/b=1.5$ and 3.
{\rm The results are plotted using filled symbols and solid lines
for prolate spheroids and using open symbols and dashe lines
for oblate spheroids.
The values of $r_{V,\,\rm cut}$ are given for the extreme models.}
Horizontal and vertical dashed lines
correspond to the average Serkowski curve
($\lambda_{\max}=0.55\,\mu$m, $K=0.92$).
Inclined dashed line corresponds to the relation
$K = 1.66  \lambda_{\max}$ (see Eq.~(\ref{k92})).}
\label{k-abt}
\end{figure}

Our modelling shows that the observed wavelength dependence of the
\is linear polarization and the correlation between $K$ and $\lambda_{\max}$
can be reproduced using imperfectly aligned silicate spheroidal grains.
Time variations of the particle size distribution are determined
by the processes of grain accretion and coagulation in the tuned model
where the coagulation threshold of silicate is removed.
The original model without tuning predicts minor variations for polarization
curves and must be rejected.

Our model has two main parameters:
the cut-off size $r_{V,\rm cut}$ in the alignment function and
grain size distribution function determined by the time of grain
processing in the molecular clouds $T$.
Increasing $r_{V,\rm cut}$ enables us to explain simultaneously the increase
of the maximum wavelength $\lambda_{\max}$ and the reduction of the width
of polarization curve (to increase parameter $K$).
As it is seen from  Fig.~\ref{k-lms}, the cut-off size must be rather
large ($r_{V,\rm cut}\ga 0.2\,\mkm$) in order to explain the
observational data with $\lambda_{\max} \ga 0.6\,\mkm$.
However, reducing  particle aspect ratio $a/b$ and replacing prolate
spheroids with oblates, one permits to decrease significantly the value
of $r_{V,\rm cut}$ (Fig.~\ref{k-abt}).\footnote{To appreciate
the influence of particle type and shape on polarization we must also examine
the behaviour of polarizing efficiency $P_{\max}/A_V$ which
anticorrelates with $\lambda_{\max}$
(Voshchinnikov et al., in preparation, see also  \citealt{voshchinnikov12}).}
Changes in $r_{V,\rm cut}$ may be attributed to the selective action of
grain alignment by the anisotropic radiation fluxes
(radiative torque alignment). This mechanism is effective if
$r_{V} \ga \lambda_{\rm eff}/2 \pi$ (\citealt{draine11}), so the trend
in Fig.~\ref{k-lms} may reflect
the systematic changes of starlight background energy distribution
(increase of the fraction of red stars with corresponding growth of
the effective wavelength $\lambda_{\rm eff}$)
from bottom left to up right corner.

Using the initial model with $T=0$~Myr and
varying  $r_{V,\rm cut}$, particle type and shape we can not reproduce a part
of observational data located at the lower left corner and middle part of
Fig.~\ref{k-lms} ($\lambda_{\max}\la 0.6\,\mkm$, $K \la 1.1$)\footnote{It
is also evident that the essential modifications of
the initial size distribution are required in order to explain
the polarimetric data of four extragalactic Type Ia Supernovae
obtained \citet{patat14}
($0.2\,\mkm \la \lambda_{\max}\la 0.4\,\mkm$, $0.8 \la K \la 1.5$).}.
In this case, to interpret observations we need to apply the tuned models
with evolution time  $T \la 30$~Myr.
These models fit the major part of observed extinction curves
except for several stars with $R_V>5$ where the models with
$T = 200 - 300$~Myr are required (see Fig.~7 in HV14).
Note that for stars with $R_V>5$  multi-wavelength polarimetric data
are not available, so they do not enter into our sample.


The restriction on the duration of accretion and coagulation in the tuned
models obtained by us ($T \la 30$~Myr) does not contradict to the lifetime
of molecular clouds found from chemical modelling (3 -- 6 Myr; \citealt{pag11})
and dynamical simulation (5 -- 25 Myr; \citealt{dobbs13}).
We should also emphasize that
the time-scales of accretion and coagulation both scale with
$\propto n_\mathrm{H}^{-1}$ (HV14), i.e. if we adopt
$n_\mathrm{H}=10^4$ cm$^{-3}$ instead of $10^3$ cm$^{-3}$,
the same size distributions are reached at ten times shorter time.


\section{Conclusions}\label{concl}

The main results of the paper can be formulated as follows.

\begin{enumerate}

\item
We applied the  grain size distributions found by \citet{hv14}
to the explanation of the \is linear polarization.
Time evolution of grain size distribution is
due to the accretion and coagulation in an interstellar cloud.
We considered the model with commonly used material parameters and
tuned model, in which coagulation of carbonaceous dust
is less efficient  and that of silicate is
more efficient with the coagulation threshold being removed.

\item
To calculate the polarization,
we used the model of homogeneous silicate and carbonaceous spheroidal
particles with different aspect ratios $a/b$ and imperfect alignment.
It was assumed that polarization is mainly produced by large silicate particles
with sizes $r_{V} \ga r_{V,\rm cut}$.
We calculated the wavelength dependence of polarization
and determined parameters of the Serkowski curve $K$ and $\lambda_{\max}$
describing  the width of the polarization curve and the
wavelength at the maximum polarization, respectively.

\item
It was found that the evolutionary effects are negligible in the original model
without tuning. This is a consequence of the insignificant evolutionary
changes of large silicate grains contributing to observed polarization.
{\rm In the tuned model, the coagulation threshold of silicate is removed,
and at $\sim (30-50)(n_\mathrm{H}/10^3 \mathrm{cm}^{-3})^{-1}$ Myr,
the maximum of polarization
displaces to the longer wavelengths and the polarization curve becomes
wider ($K$ reduces).}

\item
We compiled parameters of Serkowski curve
for a sample of 160 lines of sight and compared theory and observations.
{\rm The observed trend between $K$ and $\lambda_{\max}$ can be explained
if we use the tuned models with $T \la (30 - 40)(n_\mathrm{H}/10^3 \mathrm{cm}^{-3})^{-1}$~Myr
and different values of the cut-off size $r_{V,\rm cut}$.}
It is significant that the evolutionary effect appears in the perpendicular
direction to the effect of $r_{V,\rm cut}$ on
$K$ -- $\lambda_{\max}$ diagram.
Very narrow polarization curves ($K>1.2$) can be reproduced if we change
the type of particles (prolate/oblate) and/or to vary the particle shape
(parameter $a/b$).

\end{enumerate}

\section*{Acknowledgments}

We thank A. P. Jones for sending the refractive indexes in tabular form
and interesting discussion
and  V. B. Il'in for careful reading of manuscript.
We are grateful to M. Matsumura, the referee, for useful comments that improved
this paper.
NVV acknowledges the support from RFBR grant 13-02-00138a and
Saint-Petersburg State University grant 6.38.669.2013.
HH thanks the support from the Ministry of Science and Technology (MoST)
grant 102-2119-M-001-006-MY3.


\bsp

\label{lastpage}

\end{document}